\documentclass[prc,aps,twocolumn,showpacs,preprintnumbers,nofootinbib,amsmath,amssymb]{revtex4}
\usepackage{graphicx}
\usepackage{psfig}
\usepackage{dcolumn}

\begin{document}
\title{Collective flow of open and hidden charm in Au+Au collisions
at $\sqrt{s}$ = 200 GeV}
\author{E. L. Bratkovskaya$^1$, W. Cassing$^2$, H. St\"ocker$^1$, and N. Xu$^{3}$}
\affiliation{ \phantom{a}\\
 $^1$ Institut f\"{u}r Theoretische Physik,
      Universit\"{a}t Frankfurt, 60054 Frankfurt, Germany \\
 $^2$ Institut f\"{u}r Theoretische Physik, Universit\"{a}t Giessen, 35392 Giessen, Germany \\
 $^3$ Nuclear Science Division, Lawrence Berkeley National Laboratory, Berkeley, CA 94720, USA}

\begin{abstract}
We study the collective flow of open charm mesons and charmonia in
Au+Au collisions at $\sqrt{s}$ = 200 GeV within the
hadron-string-dynamics (HSD) transport approach. The detailed studies
show that the coupling of $D,\bar{D}$ mesons to the light hadrons leads
to comparable directed and elliptic flow as for the light mesons.  This
also holds approximately for $J/\Psi$ mesons since more than 50\% of the
final charmonia for central and mid-central collisions
stem from $D+\bar{D}$ induced reactions in the transport
calculations. The transverse momentum spectra of $D,\bar{D}$ mesons and
$J/\Psi$'s are only very moderately changed by the (pre-)hadronic
interactions in HSD which can be traced back to the collective flow
generated by elastic interactions with the light hadrons.
\end{abstract}

\maketitle

\section{Introduction}

The dynamics of  nucleus-nucleus collisions at
Relativistic-Heavy-Ion-Collider (RHIC) energies are of fundamental
interest with respect to the properties of hadronic/partonic systems at
high energy densities.  Especially the formation of a quark-gluon
plasma (QGP) and its transition to interacting hadronic matter has
motivated a large community for more than two decades
\cite{Horst86,CaMo,Cass99,QM01}.  However,  the complexity of the
dynamics has not been unraveled and the evidence  for the formation of
a QGP and/or the properties of the phase transition is much debated
\cite{SQM04}. Apart from the light and strange flavor
($u,\bar{u},d,\bar{d},s,\bar{s}$) quark physics and their hadronic
bound states in the vacuum ($p,n,\pi, K, \phi, \Lambda$ etc.), the
interest in hadrons with charm ($c, \bar{c}$) has been rising
continuously since the heavy charm quark provides an additional energy
scale, which is large compared to $\Lambda_{QCD}$.  Accordingly, the
hadronic bound states (with a $c$ or $\bar{c}$ quark) have a much
larger mass than the ordinary hadrons and one has speculated that
especially charmonia ($c\bar{c}$ bound states) might only be formed in
the very early phase of the heavy-ion collision.

In the past the charmonia $J/\Psi$, $\chi_c$, $\Psi^\prime$ have been
 discussed in context with the phase transition to the  QGP since
$c\bar{c}$ states might no longer be formed due to color screening
\cite{Satz,Satznew}. However, more recent calculations within lattice
QCD (LQCD) have shown that at least the $J/\Psi$ survives up to $\sim$
1.5 $T_c$ ($T_c \approx$ 0.17 GeV) such that the lowest $c\bar{c}$
states remain bound up to energy densities of about 5
GeV/fm$^3$ \cite{KarschJP,HatsudaJP,Karsch2}. It is presently not clear if
also the $D$ or $D^*$ mesons will survive at temperatures above $T_c$
but strong correlations between a light quark (antiquark) and a charm
antiquark (quark) are likely to persist also above $T_c$.

Moreover, it has been pointed out  (within statistical models) that at
top RHIC energies the charmonium formation from open charm and
anticharm mesons might become essential
\cite{BMS,Goren,Kost_SPS1,Kost_SPS2} and even exceed the yield from
primary nucleon-nucleon ($NN$) collisions \cite{Rafelski}. Such
concepts should work out if the early hot and dense medium created in
nucleus-nucleus collisions is very strongly interacting and an
approximate chemical equilibrium is achieved rapidly. As argued in
Refs. \cite{Cass02,BSW,Greiner} an early equilibration might be due to
multi-particle interactions either on the partonic \cite{Greiner} or
the hadronic side \cite{Cass02,BSW}.

A previous analysis within the HSD transport model \cite{Brat03} has
demonstrated that the charmonium production from open charm and
anticharm mesons becomes essential in central Au+Au collisions at RHIC.
This is in accordance with independent studies in Refs.
\cite{Rappnew,Ko}. On the other hand, these backward channels --
relative to charmonium dissociation with comoving mesons -- have been
found to be practically negligible at SPS energies. Furthermore, the
transport studies in Ref.  \cite{Kost04} have shown that chemical
equilibrium between the different charmonia $J/\Psi$, $\chi_c$,
$\Psi^\prime$ is not obtained in full phase space on the basis of
(pre-)hadronic interactions.  As pointed out in the latter study this
opens up a possibility to distinguish experimentally a statistical
freeze-out concept from a hadron/string dynamical picture.

Apart from the total and relative abundancies of charmonia and open
charm mesons also their collective properties are of interest. Here
especially the transverse momentum (or mass) spectra are expected to
provide valuable insight to the dynamics in either the very early or
late phase \cite{Dumitru,NXU,Cass01,Rapp04,Molli}.  We recall that the
transverse mass spectra at midrapidity show experimentally a 'slope'
parameter that increases ($\sim$ linearly) with the hadron mass
\cite{expslope} if the latter involve only light quarks or only a
single strange quark. Multi-strange baryons indeed show a 'slope'
parameter below the linear scaling with mass known from the lighther
hadrons \cite{XU2}. The question thus arises if the open charm mesons
will follow the trend of the light hadrons similar to kaons (involving
only a single $s$ or $\bar{s}$ quark)? Also: will the charmonia with
their substantially higher mass also show the linear trend as expected
from hydrodynamics \cite{Dumitru} or do they decouple early as expected
in hadron/string dynamical approaches due to the rather low cross
sections with light hadrons \cite{Cass01} (cf. also Ref. \cite{Rapp04})?

In this work we extend our previous studies in Refs.
\cite{Cass01,Brat03,Kost04} with respect to the collective dynamics of
$D, \bar{D}$ mesons and charmonia and concentrate in particular on the
transverse mass spectra, the in-plane flow $v_1$ and the elliptic
flow $v_2$ of these particles. The question we aim at solving is:
what is the amount of collectivity generated by (pre-)hadronic
interactions in the course of the expansion of the system? Any sincere
deviation to future measurements thus will indicate additional sources
for pressure and/or strong interactions beyond the standard
hadron/string picture.

Our study is organized as follows: After a very brief reminder
of the conceptional organization of the HSD transport approach in
Section II we will present the input for the transport calculations
with respect to the open charm and charmonium dynamics (or provide the
relevent references for details). In Section III we present our results
for the transverse momentum spectra of $D$-mesons and $J/\Psi$'s for
different centralities in Au+Au collisions at $\sqrt{s}$ = 200 GeV. In
order to see the effect of final state interactions more clearly we
will, furthermore, present ratios of momentum spectra from Au+Au
reactions relative to scaled $pp$ collisions. Section IV is devoted to
the in-plane flow $v_1$ and elliptic flow $v_2$ of open charm
mesons and charmonia as well as to the freeze-out properties of these
particles. Section V concludes this study with a summary.

\section{Basic concepts of the HSD transport approach}

We employ the HSD transport model \cite{Geiss,Cass99} for our study of
Au+Au collisions. This approach takes into account the formation and
multiple rescattering of formed hadrons as well as unformed 'leading'
pre--hadrons and thus incorporates the dominant final state
interactions. In the transport approach nucleons, $\Delta$'s,
 N$^*$(1440), N$^*$(1535), $\Lambda$, $\Sigma$ and $\Sigma^*$ hyperons,
$\Xi$'s, $\Xi^*$'s and $\Omega$'s as well as their antiparticles are
included on the baryonic side whereas the $0^-$ and $1^-$ octet states
are included in the mesonic sector. Inelastic hadron--hadron collisions
with energies above $\sqrt{s} \simeq $ 2.6 GeV are described by the
FRITIOF model \cite{LUND} (employing PYTHIA v5.5 with JETSET v7.3 for
the production and fragmentation of jets \cite{PYTHIA0}), whereas low
energy hadron--hadron collisions are modeled in line with experimental
cross sections. We stress that no explicit parton cascading or
gluon-gluon dynamics is involved in our transport calculations contrary
to e.g.~the AMPT model \cite{Ko_AMPT} or explicit parton cascades
\cite{partons}.

A systematic analysis of HSD results and experimental data for central
nucleus--nucleus collisions from 2 A$\cdot$GeV to 21.3 A$\cdot$TeV has
shown that the spectra for the 'longitudinal' rapidity distribution of
protons, pions, kaons, antikaons and hyperons are in reasonable
agreement with available data \cite{Weber02,Brat03,Brat04}.  However,
there are  problems with the dynamics in the direction transverse to
the beam. Whereas the pion transverse momentum spectra are rather well
described from lower AGS to top RHIC energies the transverse momentum
slopes of kaons/antikaons are clearly underestimated above $\sim$ 5
A$\cdot$GeV in central Au+Au collisions. In Ref. \cite{Brat04} this failure
has been attributed to a lack of pressure generation in the very early
phase of the heavy-ion collisions which also shows up in the
underestimation of the elliptic flow of charged hadrons at RHIC
energies \cite{Brat03}.

We note, that the inclusion of initial state Cronin effects gives a
substantial hardening of kaon spectra at RHIC energies, but the
slope of the pion spectra at low $m_T$ is only slightly enhanced.
In the present study  we have also included the Cronin effect,
however, found out that it has only a small impact on charm and
charmonia spectra.

Inspite of the deficiences pointed out above the overall reproduction
of the experimental hadron spectra is sufficiently realistic such that
we can proceed with open and hidden charm dynamics.

\subsection{Perturbative treatment of $D,\bar{D}$ mesons and charmonia}

The initial conditions for the production and also subsequent
propagation of $D,\bar{D}$ mesons and charmonia are incorporated in the
 HSD approach by a superposition of $pp$ collisions described via
scaled PYTHIA \cite{PYTHIA0} simulations \cite{Cass01}. For the
production and propagation of open and hidden charm hadrons  we employ
a perturbative scheme as also used in
Refs.~\cite{Cass01,Brat03,Kost04}.  Each 'perturbative' particle ($h_i$)
is produced in the transport calculation with a weight $W_i$ given by the
ratio of the actual production cross section divided by the inelastic
nucleon--nucleon cross section, e.g.
\begin{equation}
  W_i = \frac{\sigma_{NN \rightarrow h_i
    + X}(s)}{\sigma_{NN}^{\rm inelas.}(s)}.
\end{equation}
We then follow the motion of the 'perturbative' particles  within the
full background of strings/hadrons by propagating them as free
hadrons, i.e.~neglecting in--medium potentials, but compute
their collisional history with baryons and mesons or quarks and
diquarks. The actual parametrizations of the total cross sections
for $D, D^*, D_s, D_s^*$ mesons as well as their antiparticles are
given in Refs. \cite{Cass01,Brat03} and \cite{Kost04} together
with the cross sections for charmonium production in
nucleon-nucleon and pion-nucleon collisions. The only modification
introduced here -- relative to Refs. \cite{Brat03,Kost04} -- is a
change of the parametrisation for the differential production cross
section in transverse momentum. Our novel parametrisation follows
the first results from STAR \cite{STAR_v1Heta} and is given by a power law
in transverse momentum $p_T$
\begin{eqnarray}
W(p_T) \sim \left(1 + \frac{p_T}{3.5} \right)^{-8.3}\!\!\!\!.
\label{powerlow}
\end{eqnarray}
For simplicity -- and lack of further information -- we assume
that this parametrization holds for all $D, \bar{D}$ mesons as
well as charmonia.

\subsection{Cross sections involving charm hadrons}

In order to study the effect of rescattering we tentatively adopt the
following dissociation cross sections of charmonia with baryons
independent on the energy (in line with Refs.
\cite{Cass01,Brat03,Kost04}):
\begin{equation} \label{sigmacB}
\sigma_{c\bar{c}B} = 6 \ {\rm mb}; \ \sigma_{J/\Psi B} = 4 \ {\rm
mb}; \end{equation} $$\ \sigma_{\chi_c B} = 5 \ {\rm mb};
\  \sigma_{\Psi^\prime B} = 5 \ {\rm mb}.$$

\noindent
In (\ref{sigmacB}) the cross section $\sigma_{c\bar{c}B}$ stands for a
(color dipole)  pre-resonance ($c\bar{c})$ - baryon cross section,
since the $c\bar{c}$ pair produced initially cannot be identified with
a particular hadron due to the uncertainty relation in energy and time.
For the lifetime of the pre-resonance $c\bar{c}$ pair (in it's rest
frame) a value of $\tau_{c\bar{c}}$ = 0.3 fm/c is assumed following
Ref. \cite{Kharz}. This value corresponds to the mass difference of the
$\Psi^\prime$ and $J/\Psi$.

For $D, D^*, \bar{D}, \bar{D}^*$ - meson ($\pi, \eta, \rho, \omega$)
scattering we address to the calculations from Ref.  \cite{Konew,Ko}
which predict elastic cross sections in the range of 10--20 mb
depending on the size of the formfactor employed. As a guideline we use
a constant cross section of 10 mb for elastic scattering with mesons
and also baryons, although the latter might be even higher for very low
relative momenta.

The cross sections for charmonium production by open charm mesons or
the inverse comover dissociation cross sections are not well known and
the significance of these channels is discussed controversely in the
present literature (cf. Ref. \cite{Brat03} and Refs. therein).  We here
follow the simple 2-body transition model introduced in Refs.
\cite{Brat03,Kost04} with a single free parameter, i.e.  a matrix
element squared $M_0^2$, that allows to implement the backward
reactions uniquely by employing detailed balance for each individual
channel. The free matrix element has been fixed in Ref. \cite{Kost04}
in comparison to $J/\Psi$ and $\Psi^\prime$ suppression data at SPS
energies. For further details we have to refer the reader to Refs.
\cite{Brat03,Kost04} in order to avoid unnecessary repetitions.

\section{Transverse momentum spectra}

\begin{figure}[t]
\centerline{\psfig{file=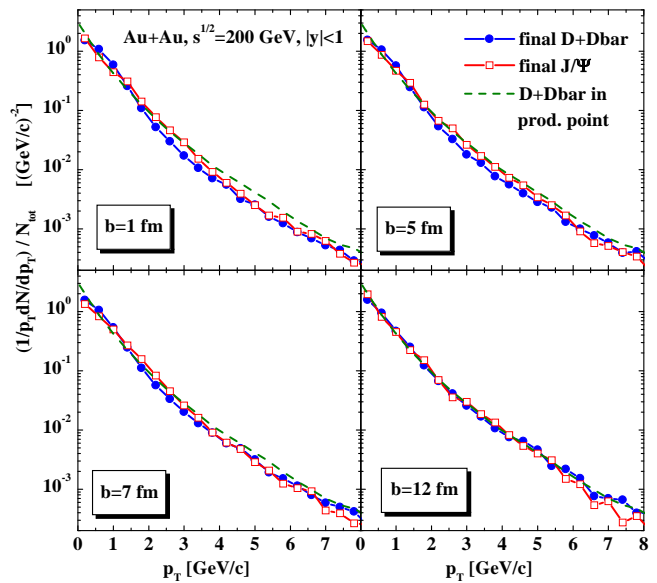,width=8.5cm}}
\vspace*{-3mm}
  \caption{(Color online)
The transverse $p_T$ spectra of the final $D+\bar D$-mesons
  (solid lines with full dots, color: blue) and $J/\Psi$
(solid lines with open squares, color: red) from Au + Au collisions at
$\sqrt{s}=200$ GeV calculated for $b=1, 5, 7$ and 12 fm at midrapidity.
The dashed lines correspond to the $p_T$ spectra of $D$-mesons at the
production point.}
  \label{pt_djp1}
\end{figure}

Since the results for charmonium suppression at SPS and RHIC energies
-- within the present approach -- have been presented in Refs.
\cite{Brat03,Kost04} we directly continue with the transverse dynamics
of open charm mesons and charmonia.  In Fig. \ref{pt_djp1} we present
the transverse $p_T$ spectra of the final $D+\bar D$-mesons  (solid
lines with full dots, color: blue) and $J/\Psi$ (solid lines with open
squares, color: red) from Au + Au collisions at $\sqrt{s}=200$ GeV
calculated for impact parameter $b=1, 5, 7$ and 12 fm at midrapidity.
The dashed lines correspond to the $p_T$ spectra of $D$-mesons at the
production point for reference, i.e. without any final state
interactions. For very peripheral Au+Au collisions we find no effects
from final state interactions whereas with increasing centrality there
is an enhancement of the spectra at low and moderate $p_T$ from Au+Au
collisions as well as a modest suppression at high $p_T$.

\begin{figure}[!]
\centerline{\psfig{file=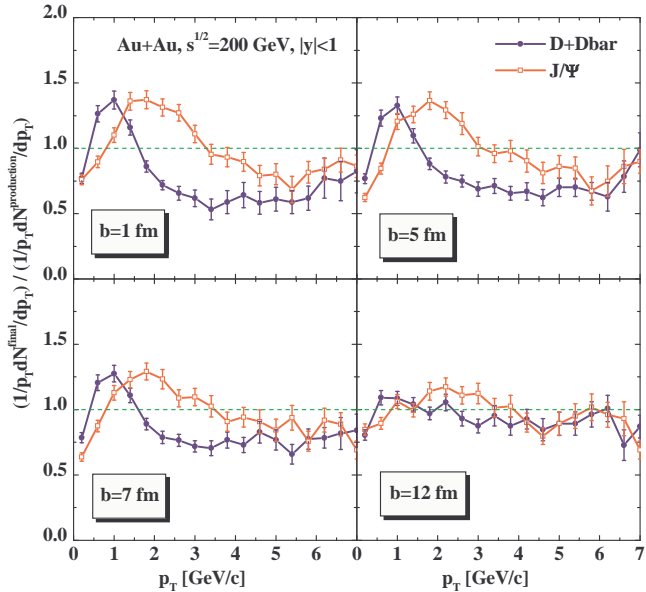,width=8.5cm}}
\vspace*{-3mm}
  \caption{(Color online)
The ratio of the final to the initial (i.e. at the
production point) transverse $p_T$ spectra of $D+\bar D$-mesons
  (solid lines with full dots, color: blue) and $J/\Psi$
(solid lines with open squares, color: red) from Au + Au collisions at
$\sqrt{s}=200$ GeV calculated for $b=1, 5, 7$ and 12 fm at midrapidity.}
  \label{pt_djp}
\end{figure}

We note that the power low parametrization (\ref{powerlow}) gives a
substantial enhancement of the high momentum tail of the open charm and
charmonia $p_T$ spectra compared to the previous parametrization used
in HSD \cite{Brat03,Kost04}.  This is an improvement in the present
approach as the new distribution provides a higher number of charm
pairs at the beginning of the collisions. This will also affect the
$J/\Psi$ yield. Further experimental data on high $p_T$ spectra in
elementary as well as heavy-ion collisions will fix the proper shape.

In order to quantify the effect of final state interactions we show in
Fig. \ref{pt_djp} the ratio of the final to the initial (i.e. at the
production point) transverse $p_T$ spectra of $D+\bar D$-mesons (solid
lines with full dots, color: blue) and $J/\Psi$ (solid lines with open
squares, color: red) from the same reaction for impact parameter $b=1,
5, 7$ and 12 fm at midrapidity.

The actual ratios in Fig. \ref{pt_djp} show an enhancement of $D,
\bar{D}$ mesons at low momenta with a maximum at $p_T \approx $ 1 GeV/c
and a relative suppression for $p_T > $ 2 GeV/c. These effects increase
with the centrality of the Au+Au collision. For the $J/\Psi$ a maximum
in the ratio shows up at $\sim 2$ GeV/c while it drops below 1 for
$p_T >$ 3--4 GeV/c.

\begin{figure}[h]
\centerline{\psfig{file=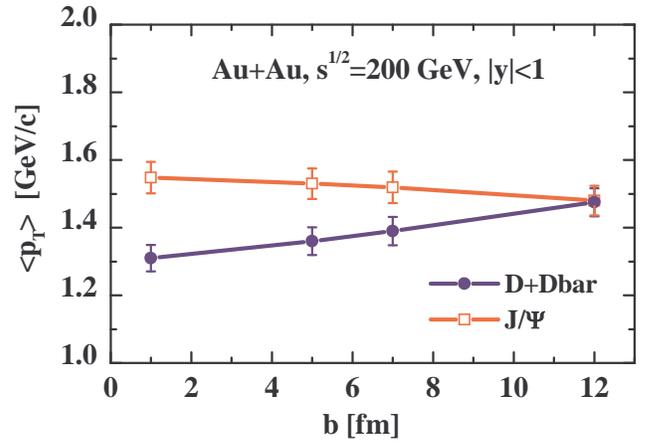,width=8.3cm}}
\vspace*{-3mm}
\caption{(Color online)
The average value of the transverse momentum  $p_T$ at midrapidity
for $D, \bar D$-mesons (solid line with full dots, color: blue) and $J/\Psi$
(solid line with open squares, color: red) from Au + Au collisions at
$\sqrt{s}=200$ GeV versus impact parameter $b$.}
  \label{ptav_b}
\end{figure}

In accordance with the ratios in Fig. 2 the averaged value of the
transverse momentum $p_T$, which is displayed  in Fig. \ref{ptav_b} at
 midrapidity for $D+\bar D$-mesons (solid line with full dots, color:
 blue) and $J/\Psi$ (solid line with open squares, color: red) from Au
+ Au collisions at $\sqrt{s}=200$ GeV,  slightly increases (decreases)
for $J/\Psi$ ($D, \bar{D}$ mesons) with decreasing impact parameter $b$.
The $D$-meson spectrum is relatively 'hard' due to the large intrinsic
charm quark mass. Through scatterings, the high $p_T$ $D$-mesons loose their
energy and develop the collective shoulder at low $p_T$. This explains the
dropping of $D$-meson $\langle p_T\rangle$ in more central Au + Au
collisions shown in Fig.  \ref{ptav_b}.  Such hadronic energy loss for
$D$-mesons will complicate the analysis of the gluon radiative energy
loss before hadronization \cite{Dima}.  These main trends should be
readily tested experimentally in the near future.

\begin{figure}[h]
\vspace*{-3mm}
\centerline{\psfig{file=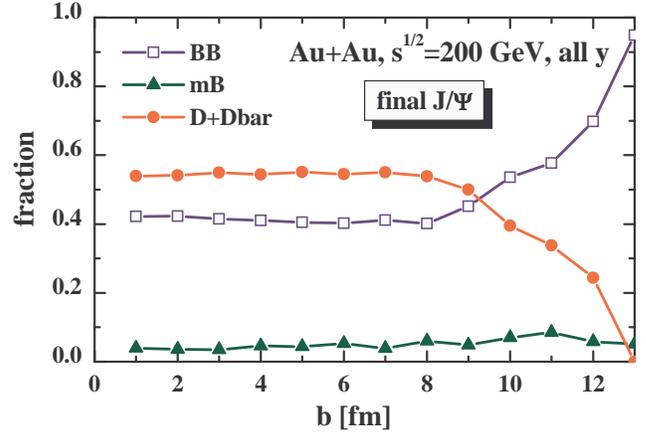,width=8.3cm}}
\vspace*{-3mm}
  \caption{(Color online)
The channel decomposition for the final $J/\Psi$ in
Au + Au collisions at $\sqrt{s}=200$ GeV within the transport approach
as a function of the impact parameter $b$.  The fraction of the final
$J/\Psi$ mesons produced in $BB$ interactions is shown by the solid
line with open squares (color: blue), the contribution from
meson-baryon ($mB$) interactions  is given in terms of the solid line
with full triangles (color: green) and the $D\bar D$ annihilation
fraction is displayed by the solid line with full circles (color: red).}
\label{cdecJP}
\end{figure}

The question comes up how to interpret the findings in Figs.
\ref{pt_djp1} to \ref{ptav_b}. We note that the maxima in the
$D+\bar{D}$ and $J/\Psi$ ratios disappear when switching off the
rescattering with mesons in the transport approach. Thus a collective
acceleration of the $D+\bar{D}$ mesons occurs dominantly via elastic
scattering with mesons. This argument does not hold at first sight for
the $J/\Psi$'s since their cross section with mesons is much lower and
no strong collective acceleration by rescattering should be expected
\cite{Cass01}. The answer comes about as follows: As known from Refs.
\cite{Brat03,Kost04} the major fraction of final $J/\Psi$'s stems from
$D+\bar{D} \rightarrow J/\Psi (\chi_c) + meson$ channels and not from
the primary production by $NN$ collisions. This finding is quantified
in Fig. \ref{cdecJP}, where the channel decomposition for the final
$J/\Psi$'s is shown as a function of the impact parameter $b$ in Au+Au
collisions at $\sqrt{s}$ = 200 GeV. Except for peripheral reactions ($b
> $ 9 fm) the $D+\bar{D}$ channel (solid line with full dots, color:
red) dominates while the production in initial baryon-baryon ($BB$)
scattering (full line with open squares, color:blue) superceeds all
other reaction channels for the most peripheral reactions. The
contribution from meson-baryon ($mB$) channels (solid line with full
triangles, color: green) is seen to be low at all centralities.

We recall that the sensitivity of $J/\Psi$ formation versus $D+\bar D$
annihilation to the formation amplitude has been studied in detail in
Refs. \cite{Brat03,Kost04}.  The amplitudes for $D+\bar D <->
J/\Psi(\chi_c, \Psi^\prime) + \rm{mesons}$ have been fitted to the
experimental data on $J/\Psi$ suppression and $\Psi'/ J\Psi$ ratios at
SPS energies and used also for the RHIC energies. This is legitimate
because the latter reactions are typical 'comover' reactions with
low/moderate relative momenta. The average relative momenta for comover
reactions do not change very much from top SPS to RHIC energies. At
RHIC energies, however, the density of produced $D$-mesons is much
higher (~17 $D+\bar D$ pairs in central collisions), and thus the
probability to annihilate to charmonia ($\sim \rho^2$) is dramatically
higher (cf.  Fig. 6 in Ref. \cite{Kost04}).  In fact, the system is
close to chemical equilibrium in central reactions such that a
sensitivity to the matrix elements is lost.

According to Fig. \ref{cdecJP} the bulk part of the $J/\Psi$'s
thus stems from $D+\bar{D}$ induced channels and their production
happens somewhat delayed in time when the $D, \bar{D}$ mesons have
already picked up collective transverse momentum. In the $D+\bar{D}
\rightarrow J/\Psi (\chi_c) + meson$ reaction then the formed $J/\Psi$
carries the collective momentum of both $D$ and $\bar{D}$ such that the
charmonium appears to be accelerated even more than the $D$-mesons. On
the other hand the $J/\Psi$'s with a high transverse momentum (in the
power law tail) partly  get dissociated with baryons of the
target/projectile or with mesons in the expansion phase of the system.
The probability to recreate a high $p_T$ charmonium by the $D+\bar{D}$
channel is lower than in the initial $NN$ channel since the invariant
energy $\sqrt{s}$ in the $D+\bar{D}$ reaction is lower by more than an
order of magnitude and exponential (thermal) $J/\Psi$ spectra become
populated. This interpretation will, furthermore, be supported by the
studies in the next Section.

\section{Directed and elliptic flow}
Apart from the transverse flow that shows up as a 'shoulder' in
the $p_T$ spectra at low momentum or a maximum in the ratio
relative to $pp$ spectra (cf. previous Section) the directed flow
\begin{equation}
\label{v1}
v_1(y,p_T) = \left\langle\frac{p_x}{p_T}\right\rangle_{y,p_T}
\end{equation}
and the elliptic flow
\begin{equation}
\label{v2}
v_2(y,p_T) = \left\langle\frac{p_x^2-p_y^2}{p_T^2}\right\rangle_{y,p_T}
\end{equation}
provide additional information on the collective currents of
hadrons in the complex reaction \cite{Stoecker}.

\begin{figure}[h]
\hspace*{-1mm}
\psfig{file=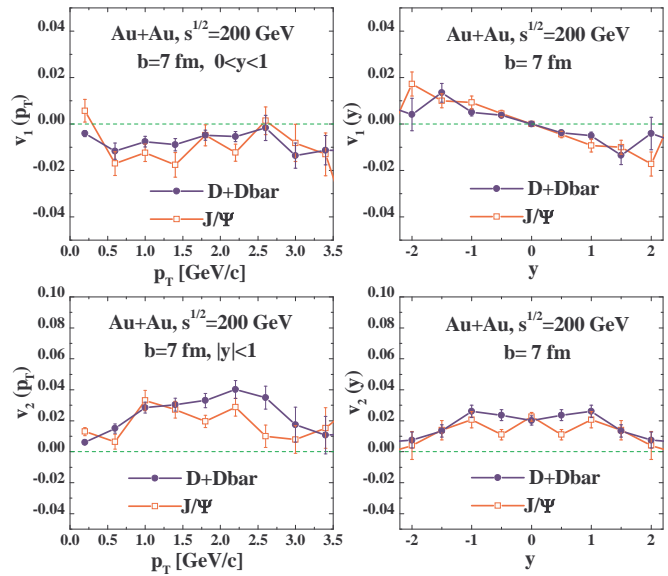,width=8.7cm}
  \caption{(Color online)
The directed flow $v_1$ (upper part) and elliptic flow $v_2$
(lower part) of $D+\bar D$-mesons (solid lines with full dots, color:
blue) and $J/\Psi$ (solid lines with open squares, color: red) from Au
 + Au collisions at $\sqrt{s}=200$ GeV for $b=7$ fm versus $p_T$ for
$0<y<1$ (left part) and rapidity $y$ (right part) integrated over $p_T$.
 }
\label{v1_v2}
\end{figure}

Fig. \ref{v1_v2} shows the HSD predictions for the directed flow $v_1$
(upper part) and elliptic flow $v_2$
(lower part) of $D+\bar D$-mesons (solid lines with full dots) and
 $J/\Psi$ (solid lines with open squares) from Au + Au collisions at
$\sqrt{s}=200$ GeV for $b=7$ fm versus $p_T$ for $0<y<1$ (left part) and
rapidity $y$ (right part) integrated over $p_T$.

The directed flow $v_1$ is negative (within statistics up to $p_T$ 3.5
GeV/c) for both $D+\bar{D}$ mesons and $J/\Psi$'s in the rapidity
interval 0 $< y <$ 1. There is a tendency that $v_1$ for $J/\Psi$'s is
larger in magnitude for 0.5 GeV/c $< p_T<$ 2 GeV/c which supports the
interpretation given in Section III that the final $J/\Psi$'s
dominantly stem from $D+\bar{D}$ channels where both open charm mesons
already have picked up some collective flow. The pronounced 'antiflow'
of open charm mesons and $J/\Psi$'s becomes visible in the r.h.s. of
Fig. \ref{v1_v2} (upper part) where the slope $d v_1/dy$ at midrapidity for
$J/\Psi$'s is also slightly larger in magnitude than that for
$D+\bar{D}$ mesons.

\begin{figure}[h]
\centerline{\psfig{file=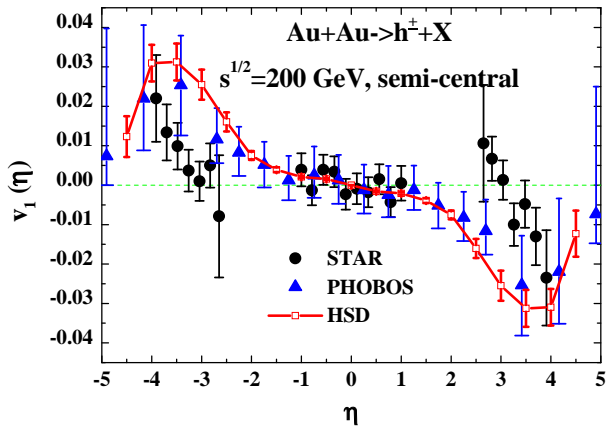,width=8cm}}
\vspace*{-3mm}
  \caption{(Color online)
The directed flow $v_1$ for charged hadrons from
 semi-central Au + Au collisions at $\sqrt{s}=200$ GeV versus
pseudorapidity  $\eta$ in comparison to the data from STAR
 \cite{STAR_v1Heta} (solid dots) and PHOBOS
\cite{PHOBOS_v1Heta} (solid triangles).}
\label{v1_etaH}
\end{figure}

The question now emerges how the 'bulk' of the light hadrons flows in
this reaction. The answer is given in Fig. \ref{v1_etaH} where the HSD
result for the directed flow $v_1$ for charged hadrons from
 semi-central Au + Au collisions at $\sqrt{s}=200$ GeV is plotted
versus pseudorapidity  $\eta$ and compared to the data from the STAR
Collaboration \cite{STAR_v1Heta} (solid dots) and the PHOBOS
Collaboration \cite{PHOBOS_v1Heta} (solid triangles). Apparently the
charged hadrons show a comparable flow $v_1(\eta)$ as the open charm
mesons and charmonia; consequently, the latter flow with the bulk of
the lighter hadrons made up from $u, d, s$ and $\bar{u}, \bar{d},
\bar{s}$ quarks. We recall that the HSD calculations for
the directed flow of charged hadrons match reasonably well with the
data from Refs. \cite{STAR_v1Heta,PHOBOS_v1Heta} for $|\eta| < 3$
(cf. Ref. \cite{bleichers} for related UrQMD calculations).

Coming back to the results for the elliptic flow $v_2$ in the lower
part of Fig. \ref{v1_v2} we notice that both $J/\Psi$'s and $D, \bar{D}$
mesons show in-plane flow since $v_2(y,p_T) > 0$. The elliptic flow of $D,
\bar{D}$ mesons is larger for 1 GeV/c $< p_T <$ 3 GeV/c than that for
$J/\Psi$'s which indicates that the $D, \bar{D}$ mesons are accelerated
earlier than the $J/\Psi$'s. The elliptic flow $v_2$ shows a maximum
around midrapity as the charged hadrons \cite{PHOBOS_v1Heta} and is also
slightly larger for $D, \bar{D}$ mesons than for $J/\Psi$'s. However,
for impact parameter $b$ = 7 fm the elliptic flow of open charm mesons
is $\leq$ 3\% whereas the elliptic flow of charged hadrons reaches up
to $\sim$ 5\% at midrapidity. We recall that the HSD calculations
underpredict the $v_2$ of charged hadrons at midrapidity by about
30-35\% \cite{Brat03}. Consequently one should expect also a larger
elliptic flow for the open charm mesons and charmonia in
experiment. We mention that in the quark coalescence model of Ref.
\cite{Rapp04} a significantly larger elliptic flow is obtained for
both $D, \bar D$ mesons and $J/\Psi$.

\begin{figure}[h]
\psfig{file=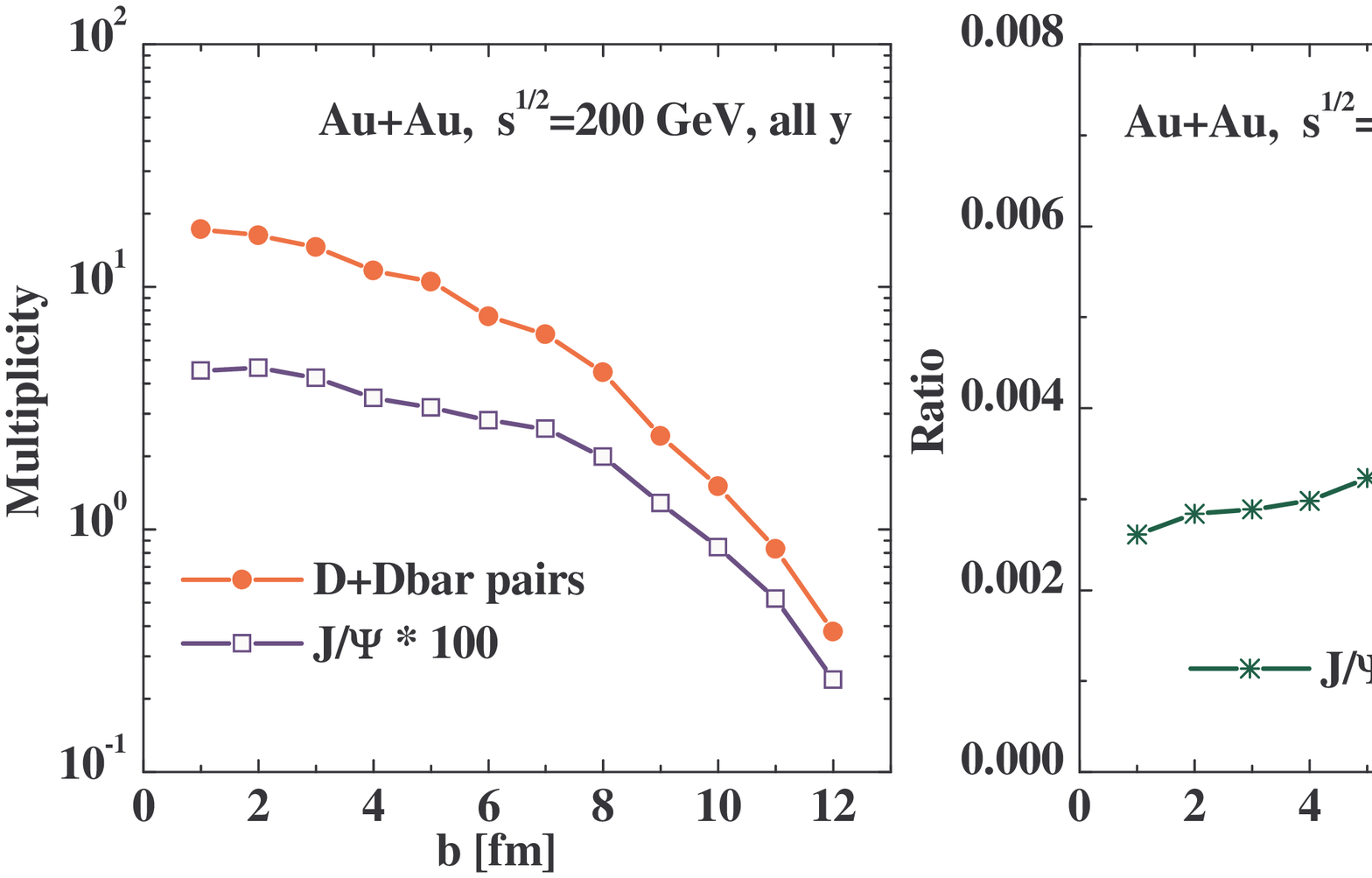,width=6.2cm}
\vspace*{-3mm}
  \caption{(Color online)
Left part: the multiplicity of $D+\bar D$ pairs (solid lines
with full dots, color: red) and  $J/\Psi$ ($\times 100$, solid lines
with open squares, color: blue) from Au + Au collisions at
$\sqrt{s}=200$ GeV integrated over rapidity   versus impact parameter
$b$.  Right part: the ratio of the multiplicity of
$J/\Psi$ over $D+\bar D$ pairs versus $b$.}
\label{rat_jpd}
\end{figure}

We now turn to the multiplicity of $D, \bar{D}$ mesons and $J/\Psi$
versus centrality in Au+Au collisions at $\sqrt{s}$ = 200 GeV. The left
part of Fig. \ref{rat_jpd} shows the multiplicity of $D+\bar D$ pairs
(solid lines with full dots, color: red) and  $J/\Psi$ (solid lines
with open squares, color: blue) from Au + Au collisions at
$\sqrt{s}=200$ GeV integrated over rapidity  versus impact parameter
$b$, whereas the right part displays the ratio of the multiplicities of
$J/\Psi$ over $D+\bar D$ pairs versus $b$. Note that the $J/\Psi$
multiplicity on the l.h.s. has been multiplied by a factor of 100.
Whereas we expect $\sim$ 17 $D+\bar{D}$ pairs in the top central
collisons the $J/\Psi$ multiplicity is only $\sim$ 5$\times 10^{-2}$,
i.e. lower by a factor of $\sim$ 300.

The r.h.s. of Fig. \ref{rat_jpd} shows the ratio of the multiplicity of
final $J/\Psi's$ over $D+\bar{D}$ pairs versus impact parameter $b$. Since the
multiplicity of $D+\bar{D}$ pairs scales directly with the number of
binary (initial) $NN$ collisions in the HSD transport approach
\cite{Cass01} this ratio gives an information about $J/\Psi$
suppression relative to binary scaled $pp$ collisions. We find -- in
accordance with the previous studies in Refs.  \cite{Brat03,Kost04} --
that the $J/\Psi$ suppression in very central collisions relative to
very peripheral reactions is about a factor of $\sim$ 2.5.
Note that the explicit shape of this ratio versus $b$ is the
result of a complex coupled-channel problem that cannot be
anticipated by simple scaling arguments. We recall that the
dominant charmonium absorption channel in our transport
calculations is the very early color-dipole dissociation with
nucleons of the target/projectile (or their baryon-like remnants).
These early dissociation reactions involve a high invariant
collision energy such that the final fragments are distributed in
a wide rapidity range which makes any recombination back to the
original color-dipole -- nucleon channel very unlikely. The
formation of charmonium states (as well as $D, \bar{D}$ mesons )
occurs later with characteristic formation times of 0.3 and 0.5
fm/c in their rest frame, respectively. Moreover, any dissociation
with mesons is delayed until the latter have formed (cf.
\cite{Brat03}). Now the dissociation of charmonia with comoving
hadrons is compensated to a large extent by the inverse $D+
\bar{D} \rightarrow $ charmonium + meson channels at top RHIC
energies. The actual time integrated reaction rates for the
forward and backward channels are shown in Fig. 5 of \cite{Kost04}
versus centrality. Since both reaction channels turn out to be
comparable in the time integrated rate for midcentral and central
collisions, the net result is not evident a priori. We recall that
the $J/\Psi$'s gain slightly from the $D+\bar{D}$ channels
whereas the $\chi_c$ and $\Psi^\prime$ states loose from the
dissociation channels with mesons (cf. Fig. 6 in \cite{Kost04}).
Now the feeddown from $\chi_c$ and $\Psi^\prime$ to $J/\Psi$'s has
to be included for the final $J/\Psi$ multiplicity which for
$Au+Au$ at top RHIC energies shows a small net suppression from
comover reactions on top of the early suppression by interactions
with projectile/target nucleons. This rather complicated coupled-channel
scenario finally leads to the ratio of $J/\Psi$ over $D+
\bar{D} $ pairs in Fig. \ref{rat_jpd}.

\begin{figure}[t]
\centerline{\psfig{file=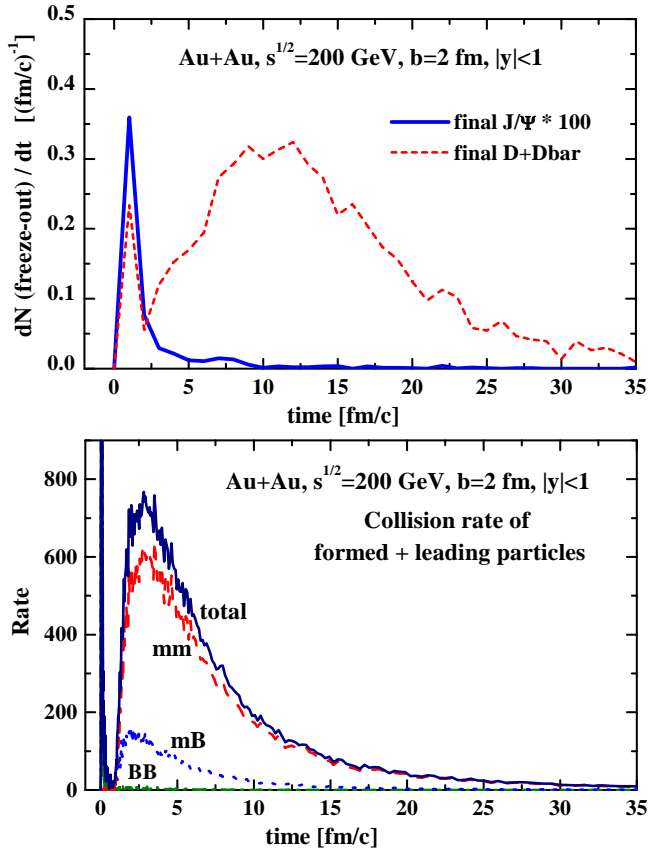,width=8.5cm}}
\vspace*{-3mm}
  \caption{(Color online)
Upper part: the distribution in time of the last interaction
for the final $J/\Psi$ mesons ($\times 100$, solid line, color: blue)
and $D+\bar D$ mesons (dashed line, color: red) from Au + Au collisions
for $b=2$ fm at $\sqrt{s}=200$ GeV ($|y| \leq$ 1).  Here the production
time is also recorded if the particles did not suffer from a further
interaction (initial peaks up to 2 fm/c).  Lower part: the time
evolution of the collision rate of formed and leading particles from
baryon-baryon ($BB$), meson-baryon ($mB$) and meson-meson ($mm$)
collisions from Au + Au collisions for $b=2$ fm at $\sqrt{s}=200$ GeV
integrated over rapidity. The time $t=0$ is defined by the contact time
of the heavy ions.}
\label{time_fz}
\end{figure}

We now turn to the freeze-out properties of $D+\bar{D}$ mesons and
$J/\Psi$'s.  The upper part of Fig. \ref{time_fz} presents the
distribution in time of the last interaction for the final $J/\Psi$
mesons ($\times 100$, solid line, color: blue) and $D+\bar D$ mesons
(dashed line, color: red) from Au + Au collisions for $b=2$ fm at
$\sqrt{s}=200$ GeV at midrapidity ($|y| \leq$ 1).  Here the production
time is also recorded if the particles did not suffer from a further
interaction (initial peaks up to 2 fm/c).  The time $t=0$ is defined by
the contact time of the heavy ions.  Whereas the $J/\Psi$ mesons
freeze-out early due to their small cross section with light hadrons
the $D+\bar D$ mesons interact for a long time and freeze-out on
average at $10-12$ fm/c.

The lower part of Fig. \ref{time_fz} shows the collision rate of formed
and leading particles from baryon-baryon ($BB$), meson-baryon ($mB$)
and meson-meson ($mm$) collisions from Au + Au collisions at $b=2$ fm
and $\sqrt{s}=200$ GeV integrated over  rapidity. Here leading diquarks
are counted as baryons ($B$) whereas leading quarks and antiquarks are
registered as mesons ($m$). The rate of $BB$ collisions shows a strong
peak during the passage time of the heavy ions ($< 0.2$ fm/c) and is
almost negligible later on. The $mB$ collision rate (apart from the
initial peak for $t <$ 0.2 fm/c) shows a sizeable contribution only
after a short time delay of $\sim$ 1 fm/c which is the scale of the
hadron formation time for low momenta in the cms. We point out that the
very early $mB$ collisions ($t <$ 0.4 fm/c) correspond to pre-hadron
interactions, where a quark from  a struck nucleon -- forming an
endpoint of the excited string -- interacts without time delay (cf.
Ref. \cite{galli} for a detailed discussion of the pre-hadron concept
in HSD). Furthermore, the early interactions of formed mesons
(for 1 fm/c $< t <$ 4 fm/c) occur dominantly in the outer transverse
region of the initial fireball where the energy density is below 1
GeV/fm$^3$.

As seen from Fig. 8 the $mm$ collision rate of formed mesons also
starts with a short delay in the order of the formation time $\tau_f$
but superceeds the $mB$ collision rate by more than a factor of 4. The
maximum in the $mm$ collision rate occurs for $t \approx 3-4$ fm/c when
the energy density in a sizeable volume drops below 1 GeV/fm$^3$ and
the system is still very dense. We point out, furthermore, that the
'hadronic burning' by meson-baryon and meson-meson collisions persists
up to rather long times since bunches of close-by hadrons may interact
almost continuously if their relative momenta are low.

When comparing the upper part of Fig. \ref{time_fz} with its lower part
one notices that the maximum in the $mm$ collision rate occurs much
earlier than the average freeze-out of $D+\bar{D}$ mesons (at 9--12
fm/c).  This further supports our conjecture in the end of Section III
that the $D+\bar{D}$ mesons suffer a couple of interactions with the
light mesons and pick up collective flow before they freeze-out rather
late on a time scale that is only slightly shorter (but comparable) to
the freeze-out time for light hadrons if one discards the very late
decay of resonances (e.g. $\omega \rightarrow 3 \pi's$, $\Delta
\rightarrow N \pi$, $K^* \rightarrow K \pi$ etc.).

\section{Summary}

In this work we have extended our previous investigations within the HSD
transport approach in Refs. \cite{Cass01,Brat03,Kost04} with respect
to the collective dynamics of $D, \bar{D}$ mesons and charmonia.  Our
detailed studies of Au+Au collisions at $\sqrt{s}$ = 200 GeV have
shown that:

\begin{itemize}

\item{Pre-hadronic and hadronic interactions generate a transverse
collective flow of $D, \bar{D}$ mesons and charmonia that shows up as
a 'shoulder' in the low momentum spectra or as a maximum in the ratio
relative to scaled $pp$ collisions.}

\item{The high $p_T$ power law tail of the spectra is only moderately
suppressed.}

\item{The directed flow of $D, \bar{D}$ mesons and charmonia is
comparable to that of the light charged hadrons.}

\item{The elliptic flow of $D, \bar{D}$ mesons and charmonia is
smaller than that of the light hadrons.}

\item{The $D, \bar{D}$ mesons freeze out on average at $\sim 9-12$
fm/c which is shorter than the freeze-out time for light
hadrons (when neglecting explicit resonance decays).}

\end{itemize}

Whereas the collective acceleration of the $D+\bar{D}$ mesons via
elastic scattering with mesons in the expansion phase of the
'fireball' can be well understood on the basis of the rather large
cross section of $D, \bar{D}$ mesons with nucleons and mesons
\cite{Konew,Ko} -- in view of the light quark/antiquark content of
these states -- the collective dynamics of the $J/\Psi$'s cannot be
explained in this way since the $J/\Psi$ cross section with mesons is
much lower. However, the dominant fraction ( $> $ 50\%) of final
$J/\Psi$'s in central and mid-central reactions stems from $D+\bar{D}
\rightarrow J/\Psi (\chi_c) + meson$ channels and to a lower extent
from the primary production by $NN$ collisions (cf. Fig. 4).  Thus the
production of the final charmonia happens delayed in time when the $D,
\bar{D}$ mesons have already picked up collective flow from
interactions with light (and fast) hadrons. Furthermore, in the
$D+\bar{D} \rightarrow J/\Psi (\chi_c) + meson$ reaction the formed
$J/\Psi$ carries the collective momentum of both $D$ and $\bar{D}$
such that the charmonium appears to be accelerated even more than the
$D$-mesons (cf. Figs. 2 and 5). On the other hand the elliptic flow of
$D, \bar{D}$ mesons is slightly larger than that of $J/\Psi$'s which
indicates on average an earlier production of the open charm mesons.

The differential spectra, ratios as well as the differential flow
analysis for $v_1(y,p_T)$ and $v_2(y,p_T)$ for open charm hadrons and
charmonia is expected to be proved/disproved in the near future by the
RHIC experiments. Our analysis and interpretation of the results has
also paved the way to extract relevant time scales and possibly
interaction rates or freeze-out times. Sensible deviations from our
predictions will points towards a dynamical origin that is not
included in our present calculations and should be addressed to
explicit partonic interactions in a possibly colored medium.

\section*{Acknowledgement}

The authors like to thank M. Bleicher, P. Braun-Munzinger, A. P.
Kostyuk, A.  Mishra, J. Schaffner-Bielich and L. Tolos for valuable
discussions. E.L.B. was supported by DFG and GSI. N.X. was supported by
the U.S. Department of Energy under Contract No. DE-AC03-76SF00098.



\begin{thebibliography}{99}
\bibitem{Horst86}
    H. St\"ocker and W. Greiner, Phys. Rep. {\bf 137}, 277 (1986).
\bibitem{CaMo}
    W. Cassing and U. Mosel, Prog. Part. Nucl. Phys. {\bf 25}, 235 (1990).
\bibitem{Cass99}
    W. Cassing and E. L. Bratkovskaya, Phys. Rep. {\bf 308}, 65 (1999).
\bibitem{QM01}
    {\it Quark Matter 2002}, Nucl. Phys. A {\bf 715}, 1 (2003);
    {\it Quark Matter 2004}, J. Phys. G {\bf 30}, S633 (2004).
\bibitem{SQM04}
    {\it Strange Quark Matter 2003}, J. Phys. G {\bf 30}, 1 (2004).
\bibitem{Satz}
    T. Matsui and H. Satz,  Phys. Lett. B {\bf 178}, 416 (1986).
\bibitem{Satznew}
    H. Satz, Rep. Progr. Phys. {\bf 63}, 1511 (2000).
\bibitem{KarschJP}
    S. Datta, F. Karsch, P. Petreczky, and I. Wetzorke,
    J. Phys. G {\bf 30}, S1347 (2004).
\bibitem{HatsudaJP}
    M. Asakawa and T. Hatsuda, J. Phys. G {\bf 30}, S1337 (2004).
\bibitem{Karsch2}
    F. Karsch, J. Phys. G {\bf 30}, S889 (2004).
\bibitem{BMS}
    P. Braun-Munzinger and J. Stachel,
    Phys. Lett. B {\bf 490}, 196 (2000);
    Nucl. Phys. A {\bf 690}, 119c (2001).
\bibitem{Goren}
    M. I. Gorenstein, A. P. Kostyuk, H. St\"ocker, and W. Greiner,
    Phys. Lett. B {\bf 509}, 277 (2001);
    J. Phys. G {\bf 27}, L47 (2001).
\bibitem{Kost_SPS1}
     A. P. Kostyuk, M. I. Gorenstein, H. St\"ocker, and W. Greiner,
    Phys. Lett. B {\bf 531},  195 (2002).
\bibitem{Kost_SPS2}
    A. P. Kostyuk, M. I. Gorenstein, H. St\"ocker, and W. Greiner,
    J. Phys. G {\bf 28},  2297 (2002).
\bibitem{Rafelski}
    R. L. Thews, M. Schroedter, and J. Rafelski,
    Phys. Rev. C {\bf 63}, 054905 (2001).
\bibitem{Cass02}
    W. Cassing, Nucl. Phys. A {\bf 700}, 618 (2002).
\bibitem{BSW}
    P. Braun-Munzinger, J. Stachel and C. Wetterich,
    Phys. Lett. B {\bf 596}, 61 (2004).
\bibitem{Greiner}
    Z. Xu and C. Greiner, hep-ph/0406278.
\bibitem{Brat03}
    E. L. Bratkovskaya, W. Cassing, and H. St\"ocker,
    Phys. Rev. C {\bf 67}, 054905  (2003).
\bibitem{Rappnew}
    L. Grandchamp and R. Rapp, Phys. Lett. B {\bf 523}, 60 (2001);
    Nucl. Phys. A {\bf 709}, 415 (2002).
\bibitem{Ko}
    Z. Lin and  C. M. Ko, J. Phys. G {\bf 27}, 617 (2001);
    Z. W. Lin and C. M. Ko, Phys. Rev. C {\bf 65}, 034904 (2002).
\bibitem{Kost04}
     E. L. Bratkovskaya, A. P. Kostyuk, W. Cassing, and H.~St\"ocker,
      Phys. Rev.  C {\bf 69},  054903  (2004).
\bibitem{Dumitru}
    A. Dumitru and C. Spieles, Phys. Lett. B {\bf 446}, 326 (1999).
\bibitem{NXU}
    N. Xu, Prog. Part. Nucl. Phys. {\bf 53}, 165 (2004)
    and references therein.
\bibitem{Rapp04} 
    V. Greco, C.M. Ko, and R. Rapp, Phys. Lett. B {\bf 595}, 202 (2004).
\bibitem{Molli}
    Z. W. Lin and D. Molnar,
    Phys. Rev. C {\bf 68}, 044901 (2003).
\bibitem{Cass01}
    W. Cassing, E. L. Bratkovskaya, and A. Sibirtsev,
        Nucl. Phys. A {\bf 691}, 753 (2001).
\bibitem{expslope} 
     I. G. Bearden {\it et al.}, (NA44 collaboration),
    Phys. Rev. Lett. {\bf 78}, 2080 (1997).
\bibitem{XU2}
     H. van Hecke, H. Sorge, and N. Xu,
     Phys. Rev. Lett. {\bf 81}, 5764 (1998).
\bibitem{Geiss}
    J. Geiss, W. Cassing,  and C. Greiner,
    Nucl. Phys. A {\bf 644}, 107 (1998).
\bibitem{LUND}
    H. Pi, Comp. Phys. Commun. {\bf 71}, 173 (1992).
\bibitem{PYTHIA0}
        H.-U. Bengtsson and T. Sj\"ostrand,
        Comp. Phys. Commun. {\bf 46}, 43 (1987).
\bibitem{Ko_AMPT}
    Z. W. Lin {\it et al.}, Nucl. Phys. A {\bf 698}, 375 (2002).
\bibitem{partons}  
    S. A. Bass, B. M\"uller, and D.K. Srivastava,
    J. Phys. G {\bf 30}, S1283 (2004).
\bibitem{Weber02}
    H. Weber, E. L. Bratkovskaya, W. Cassing, and H. St\"ocker,
    Phys. Rev. C {\bf 67}, 014904 (2003).
\bibitem{Brat04}
    E. L. Bratkovskaya, M. Bleicher, M. Reiter {\it et al.},
    Phys. Rev. C {\bf 69}, 054907 (2004).
\bibitem{STAR_v1Heta}
    An Tai {\it et al.}, STAR Collaboration,
    J. Phys. G {\bf 30}, S809 (2004).
\bibitem{Kharz}
    D. Kharzeev and R. L. Thews, Phys. Rev. C {\bf 60}, 041901 (1999).
\bibitem{Konew}
    Z. Lin and  C. M. Ko, Phys. Rev. C {\bf 62}, 034903 (2000).
\bibitem{Dima}
    Y.L. Dokshitzer and D.E. Kharzeev, Phys. Lett. B {\bf 519}, 199 (2001);
    M. Djordjevic and M. Gyulassy, J. Phys. G {\bf 30}, S1183 (2004).
\bibitem{Stoecker}
    H. St\"ocker, nucl-th/0406018.
\bibitem{PHOBOS_v1Heta}
    S. Manly {\it et al.}, PHOBOS Collaboration, nucl-ex/0405029.
\bibitem{bleichers}
    M. Bleicher and H. St\"ocker, Phys. Lett. B {\bf 526}, 309 (2002).
\bibitem{galli}
    W. Cassing, K. Gallmeister, and C. Greiner,
    Nucl. Phys. A {\bf 735}, 277 (2004).
\end{thebibliography}
\end{document}